\documentclass[%
 reprint,
 amsmath,amssymb,
 aps,
pra,
]{revtex4-2}

\usepackage{booktabs}

\usepackage{graphicx}% Include figure files
\usepackage{dcolumn}% Align table columns on decimal point
\usepackage{bm}% bold math
\usepackage{soul}%To highlight text
\usepackage{multirow} %To combine columns in table
\usepackage{xcolor}
\usepackage{comment}
\usepackage{physics}
\usepackage{natbib}

\bibstyle{article}

\begin{document}

%\preprint{APS/123-QED}
%%%%%%%%%%%%%%%%%%%%%%%%%%%%%%%%%%%%%%%%%%%%%%%%%%%%%%%%%%%%%%%%%%%%%%%%%%%%%%%%%%%%%%%%%%%%%%%%%%%%%%
\newcommand{\omegaone}{$\omega_1$}
\newcommand{\omegatwo}{$\omega_2$}
\newcommand{\tomegatwo}{$\Delta t_{\omega_2}$}
\newcommand{\trf}{$\Delta t_{\omega_{\rm{rf}} }$}
\newcommand{\tramp}{$\Delta t_{\text{ramp}}$}
\newcommand{\wrf}{$\rm W_{\text{rf}}$}
\newcommand{\trminwrf}{$\Delta t_{\text{ramp}}-\rm W_{\text{rf}}$}
\newcommand{\vpp}{$V_{pp}$}
%%%%%%%%%%%%%%%%%%%%%%%%%%%%%%%%%%%%%%%%%%%%%%%%%%%%%%%%%%%%%%%%%%%%%%%%%%%%%%%%%%%%%%%%%%%%%%%%%%%%%

\title{Radio frequency field-induced electron mobility in an ultracold plasma state of arrested relaxation}

\author{R. Wang}
\affiliation{Department of Physics \& Astronomy, University of British Columbia, Vancouver, BC V6T 1Z3, Canada}
\author{M. Aghigh} 
\affiliation{Department of Chemistry, University of British Columbia, Vancouver, BC V6T 1Z3, Canada}
\author{K.  L. Marroqu\'in} 
\affiliation{Department of Chemistry, University of British Columbia, Vancouver, BC V6T 1Z3, Canada}
\author{K. M. Grant} 
\affiliation{Department of Chemistry, University of British Columbia, Vancouver, BC V6T 1Z3, Canada}
\author{J. Sous}  
\affiliation{Department of Physics, Columbia University, New York, NY 10027, USA}
\author{F. B. V. Martins} 
 \affiliation{Laboratory of Physical Chemistry, ETH Zurich, CH-8093 Zurich, Switzerland}
\author{J. S. Keller}
 \affiliation {Department of Chemistry, Kenyon College, Gambier, Ohio 43022 USA}
\author{E. R. Grant}
\email[Author to whom correspondence should be addressed. Electronic mail:  ]
{edgrant@chem.ubc.ca}
\affiliation{Department of Chemistry, University of British Columbia, Vancouver, BC V6T 1Z3, Canada}
\affiliation{Department of Physics \& Astronomy, University of British Columbia, Vancouver, BC V6T 1Z3, Canada}

\begin{abstract}

Penning ionization releases electrons in a state-selected Rydberg gas of nitric oxide entrained in a supersonic molecular beam.  Subsequent processes of electron impact avalanche, bifurcation and quench form a strongly coupled, spatially correlated ultracold plasma of NO$^+$ ions and electrons that exhibits characteristics of self-organized criticality. This plasma contains a residue of nitric oxide Rydberg molecules.  A conventional fluid dynamics of ion-electron-Rydberg quasi-equilibrium predicts rapid decay to neutral atoms.  Instead, the NO plasma endures for a millisecond or more, suggesting that quenched disorder creates a state of suppressed electron mobility.  Supporting this proposition, a 60 MHz radio frequency field with a peak-to-peak amplitude less than 1 V cm$^{-1}$ acts dramatically to mobilize electrons, causing the plasma to dissipate by dissociative recombination and Rydberg predissociation.  An evident density dependence shows that this effect relies on collisions, giving weight to the idea of arrested relaxation as a cooperative property of the ensemble. 

\end{abstract}

\maketitle

\section{Introduction}

At a given temperature and density, the properties of its most stable phase determine the natural state of a quantum many-body system.  For instance, a lattice of cations in a sea of delocalized electrons naturally affords a metal with ductility and a band structure that supports electrical conductivity.  Covalent bonds and intermolecular forces organize the atoms in a molecular solid to form a crystal.  But, out of equilibrium, in a strongly correlated material, cooperative properties can impose characteristics quite apart from the long-time limit of Boltzmann statistical mechanics.  Such conditions can produce transient states that persist for a range of times.  

In other words, strong correlations in out-of-equilibrium many-body systems offer paths to collective behaviour that cannot occur as a natural property of a ground state, or an excited state in its statistical limit \cite{takei2016direct,bruder2019delocalized,mizoguchi2019ultrafast}. Thus, for example, correlations in a crystal can cause a particular electromagnetic transition to form a transient high-temperature superconducting phase \cite{Fausti2011,Mitrano2016}.  More generally, an excited system can show signs of prethermalization \cite{Demler2012}, or undergo a nonequilibrium phase transition in which physical quantities quench to become nonanalytic functions of time \cite{kadau2016observing,Eckardt2017,cooper2019}.  

In particular cases, a driving force balanced by dissipation can organize the microscopic behaviour of a strongly coupled dynamical system to exhibit spatial and temporal scale invariance  \cite{bak1987}.  This dynamical property, which parallels the critical point of a conventional phase transition, does not depend on external control parameters, but rather approaches a critical attractor in a process of self-organization -- a propagating avalanche with large spatiotemporal correlations known as self-organized criticality (SOC) \cite{turcotte1999,aschwanden2013}.  Atomic-scale self-organized criticality propagates much like large-scale macroscopic phenomena such as earthquakes \cite{bak1989} and forest fires \cite{drossel1992}.  

Some of the earliest work on the phenomenon of self-organized criticality focused on resilient, non-local properties of cold plasmas \cite{newman1996,alex2017}, and  plasma structures confined magnetically or by pressure-gradient turbulence \cite{carreras1996}, as manifestations of SOC avalanche and transport dynamics.  Recent experiments have found signs of self-organizing criticality in ultracold and room-temperature atomic Rydberg gases \cite{helmrich2020signatures,ding2020phase}.  For some time, we have observed elements of self-organization in the properties of the nitric oxide molecular ultracold plasma \cite{Plasma_prl,Sadeghi:2014,schulz2016evolution,Haenel2017,mmWave1}. 

In our experiment, double-resonant excitation of ground-state nitric oxide cooled in a supersonic molecular beam forms a state-selected Rydberg gas, which undergoes spontaneous Penning ionization.  This triggers an electron impact avalanche that splits the cross-beam ellipsoid of NO$^+$ ions and Rydberg molecules \cite{Schulz-Weiling2016}.  Ambipolar expansion quenches electron temperature of this ultracold plasma.  Long-range resonant charge transfer from ballistic ions to frozen Rydberg molecules quenches the ion-Rydberg relative velocity \cite{Haenel.2018}. This sequence of steps gives rise to a remarkable mechanics of self-assembly that forms a gas of canonical density, characteristic of self-organized criticality.  In this process, the plasma expends the kinetic energy of initially formed hot electrons to separate plasma volumes in the laboratory frame \cite{review}. These dynamics sequester energy in a reservoir of mass transport, starting a process that anneals recoiling volumes to form a scale-invariant ultracold glass of strongly coupled ions and electrons. 

This ultracold plasma state of arrested relaxation persists for a millisecond or longer, directly signalling the absence of dissociative recombination collisions of electrons and ions.  Evident as well is a suppressed state of electron-Rydberg collisions.  The long-lived ultracold plasma traps an adventitious population of Rydberg molecules with principal quantum number, $n$, distributed about $n_0$, the value selected by double-resonant preparation of the Rydberg gas.  These electronically excited molecules predissociate at a rate sensitively regulated by $n$ and $\ell$, the orbital angular momentum quantum number of the Rydberg electron.  In the collision-free environment, molecules of low $\ell$ rapidly predissociate, leaving a long-lived, high-$\ell$ residue.  The persistence of this residue acts as responsive sensor of electron mobility.  Any process that stimulates electron collisions scrambles $\ell$, restarting predissociation.  

The sequence of events that leads to self-organized criticality in the bifurcated nitric oxide ultracold plasma forms a quantum mechanical state of ultracold electrons and spatially correlated, locally stationary NO$^+$ ions.  We have argued that such a system of coupled dipoles supports a strength of dipole-dipole flip-flop interaction that compares with the global van der Waals energy to a degree that predicts a systematic destruction of transport \cite{Sous2018,Sous2019}.   

Here, we offer direct experimental support for this picture.  A radio frequency field, applied to a plasma in a state of arrested relaxation, acts to mobilize electrons, restarting dissociative recombination and Rydberg predissociation.  The fraction of molecules consumed scales with density, proving that this process relies on collisions:  The field must operate on the plasma volume as a whole, as opposed to a perturbation of individual molecules.  This offers support for the idea of arrested relaxation as a cooperative property of the ensemble in a state with characteristics of many-body localization.

%%%%%%%%%%%%%%%%%%%%%%%%%%%%%%%%%%%%%%%%%%%%%%%%%%%%%%%%%%%%%%%%%%%%%%%%%%%%%%%%%%%%%%%%%%%%%%%%%%%%%%%%%%%%%%%%%%%%%%%%%%%%%%%%%%%%%%%%

\section{Experimental methods}

\subsection{Supersonic molecular beam ultracold plasma spectrometer}

A pulsed free-jet of NO seeded 1:10 in helium expands through 0.5 mm nozzle from a stagnation pressure of 5 bar and propagates 35 mm to enter an experimental chamber through a 1 mm diameter skimmer, as diagrammed in Figure \ref{fig:experiment}.  The collimated supersonic molecular beam travels 70 mm to transit the entrance aperture of a first field plate (G$_1$).  A second grid G$_2$, held at an externally adjustable distance from G$_1$, defines a flight path of controlled field. 

Co-propagating unfocussed Q-switched Nd:YAG pumped dye laser pulses, $\omega_1$ and $\omega_2$, cross the molecular beam 6 mm beyond G$_1$.   A spatial filter collimates $\omega_1$ to propagate as a cylindrical Gaussian.  The 1 mm (fwhm) diameter of $\omega_1$ defines an ellipsoidal illuminated volume in the 3 mm diameter molecular beam.  In this volume, double resonant excitation creates a gas of state-selected high-Rydberg molecules with initial principal quantum number, $n_0$.  

\subsection{Selective field ionization}

During \omegaone + \omegatwo\ laser excitation, with G$_1$ held to ground, an adjustment G$_2$ over a range of $\pm 100$ mV serves to define a field-free region between G$_1$ and G$_2$.  At a predetermined time after $\omega_2$, a -3 kV square-wave pulse from a Behlke high-voltage switch, coupled to G$_1$ through a $10 ~ {\rm k\Omega}$ resistor, forms an electron-forward-bias voltage ramp that rises at a rate of $\sim 0.8$ V/cm ns.  We precisely fit a polynomial function to the leading edge of this voltage pulse that transforms the time-dependent electron signal waveform to an ionization spectrum as a function of field in V/cm.  

\subsection{Radio-frequency electric field}

The experiment uses a Tektronix AWG 7102 10 GS/s arbitrary waveform generator to produce radio frequency (RF) pulses of selected frequency with an amplitude from 0.1 to 1.5 V peak-to-peak.  A LabVIEW user interface selects the frequency and amplitude, with controlled delay and pulse duration.  Measurements described below use a frequency of 60 MHz, an amplitude of 400 mV cm$^{-1}$ and a pulse duration of 250 ns (W$_{\rm rf}$), triggered at a time, $\Delta t_{\omega_{\rm rf}}$ between $t_{\omega_2}$ and $t_{ramp}$.  The programmed output of the waveform generator connects to G$_2$. 

\subsection{Pulse sequence and evolution of $n_0$ Rydberg density as observed in a typical RF-SFI experiment}

Figure \ref{fig:experiment} diagrams the sequence of optical and electronic pulses used in carrying out a typical RF-SFI experiment.  The horizontal axis indicates the elapsed time after \omegatwo. Waveforms \omegaone \ and \omegatwo, represent the first and second laser pulses, and \tomegatwo \ refers to the time delay between them.  We apply an RF pulse of width W$_{\rm rf}$, with a time delay of \trf \ after the second laser pulse.  A voltage ramp that begins at a time $\Delta t_{\rm ramp}$ after $\omega_2$ selectively field ionizes the excited molecular system.  In this study, \wrf \ has a duration of 250 ns for all experiments.  We vary the intervals, \trf, \tomegatwo, and $\Delta t_{\rm ramp}$, according to the requirements of the measurements described above.

\section{Results}

\subsection{Field-free evolution of the nitric oxide molecular ultracold plasma}

\subsubsection{Density distribution of the plasma ellipsoid}

\begin{figure*}
    \centering
    \includegraphics[width= .7 \textwidth]{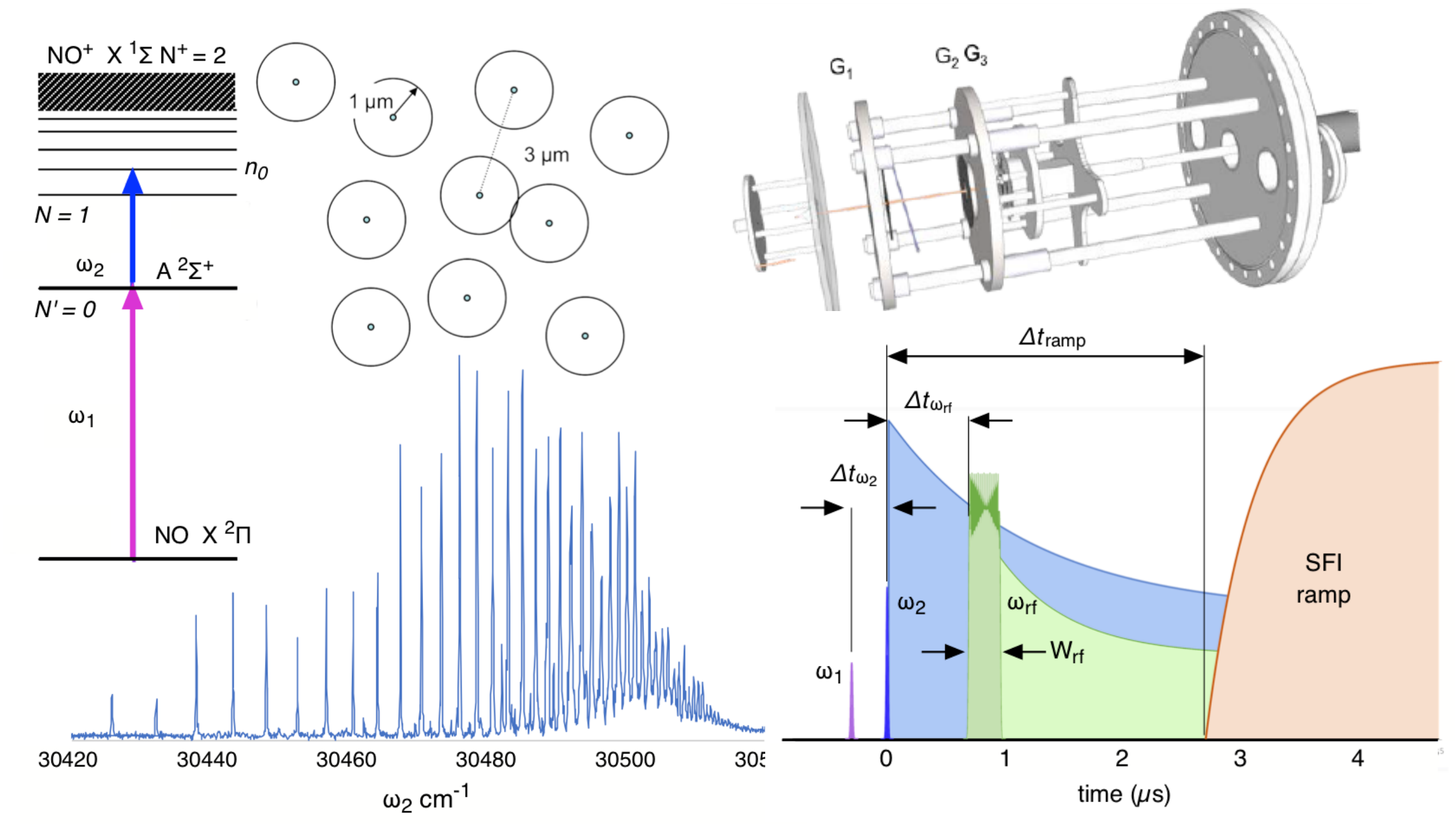}
    \caption{(left) Diagram illustrating the double-resonant excitation of a molecular Rydberg gas of nitric oxide, and the conditions leading to Penning ionization and avalanche to an ultracold plasma.  The atomic-like plasma-excitation spectrum consists exclusively of $N=1$ $n_0f(2)$ Rydberg state resonances converging to the $N^+=2$ rotational limit of NO$^+$.  (top right) Co-propagating laser beams, \omegaone \ and \omegatwo \ cross a molecular beam of nitric oxide between entrance aperture G$_1$ and grid G$_2$ of a differentially-pumped vacuum chamber. (bottom right) Schematic diagram showing the sequence of pluses in in the RF-SFI experiment. The second step of laser excitation follows the first with a specified delay, \tomegatwo.  A radio-frequency field with an adjustable peak-to-peak amplitude interacts with this ensemble, either as a CW field or as a pulse with a duration \wrf \ applied at a time, \trf, \ after \omegatwo. An electric field ramp from 0 to 350 V/cm with a rise-time of 1 $\mu$s 0.8 V/ns, applied after a delay, $\Delta t_{\rm ramp}$ \ following \omegatwo, ionizes the excited molecular system. Shaded regions represent the surviving fractions of the $n_0$ Rydberg molecules to a form a residual fraction of long-lived molecules in the absence (blue) and after the presence of a 60 MHz radio frequency field (here represented by the pulse in green). }
    \label{fig:experiment}
\end{figure*}

Double-resonant ($\omega_1 + \omega_2$) excitation of nitric oxide in a molecular beam forms a quantum-state selected ellipsoidal volume of Rydberg gas.  Figure \ref{fig:experiment} diagrams these steps of double resonant excitation from the electronic ground state X $^2\Pi_{1/2}~ (v''=0,~N''=0)$ to the intermediate A $^2\Sigma^+~(v'=0,~N'=0)$ state, and then to the selected Rydberg state.  The choice of A-state $N'=0$ allows only final Rydberg states of total angular momentum neglecting spin of $N=1$.  

The present illumination conditions yield Gaussian widths ($\sigma_x:\sigma_{y,z}$) in a ratio of approximately 3:1, where $z$ defines the propagation direction of the molecular beam and $x$ denotes the laser axis.  At $t=0$, the peak density of the Rydberg gas depends on the intensity of $\omega_1$, the laser that drives the first step of double resonance from X $^2\Pi_{1/2} ~ N''=1$ to A $^2\Sigma^+ ~ N'=0$ ($\omega_1$).  At a fixed intensity of $\omega_1$, this density falls exponentially with the time delay between the laser pulses promoting first and second steps.  The density of the Rydberg gas varies in space along the axes of this illuminated ellipsoid according to the cylindrical Gaussian photon density of $\omega_1$ crossed by the wider cylindrical Gaussian nitric oxide density of the molecular beam.  

Even in a Rydberg gas of comparatively low density, some fraction of initially excited $n_0$ molecules populate the leading tail of the nearest-neighbour distance distribution, separated by an orbital diameter or less.  Thus, for an an initial $n_0=50$ Rydberg gas density of  10$^{10}$ cm$^{-1}$, the orbital radius is about 1 $\mu$m while the average spacing between Rydberg molecules is 3 $\mu$m.  But, a good portion of the nearest-neighbour distance distribution falls within 1 $\mu$m.  These closely spaced pairs interact by Penning ionization to form prompt electrons, which seed the avalanche to ultracold plasma \cite{Sadeghi:2014}.  Among the $N=1$ high-Rydberg states accessible to $\omega_2$ excitation, only those in the $n_0f(2)$ series converging to the nitric oxide cation state, NO$^+$ X $^1\Sigma^+$ $N^+=2$ have sufficient lifetime to sustain this avalanche, as illustrated by the excitation spectrum in Figure  \ref{fig:experiment}. 

We adjust the pulse energy of the second laser to saturate the $\omega_2$ transition.  Under such conditions, the initial density of the Rydberg gas depends entirely on the instantaneous number of intermediate A $^2\Sigma^+$ molecules.  The experiment uses two means to regulate this quantity.  Varying the $\omega_1$ pulse energies from 2 to 6 $\mu$J increases the intermediate state density linearly to a degree that approaches saturation.  This population decays with a  radiative lifetime of 192 ns \cite{Settersten2009Radiative}, and thus, a delay of the $\omega_2$ laser pulse with respect to $\omega_1$ offers a precise means to vary the intermediate state density available to form a Rydberg gas at any chosen $\omega_1$ pulse energy.   Control of $\omega_1$ pulse energy and $\omega_1-\omega_2$ delay yields initial Rydberg gas peak densities from 10$^{10}$ to 10$^{12}$ cm$^{-3}$.  

To build realistic simulations of the collisional rate processes that give rise to the avalanche, we numerically  approximate this ellipsoid by a system of 100 concentric shells  \cite{MSW_tutorial,Haenel.2018}.  For example, in a laser-crossed molecular beam Rydberg gas of nitric oxide with an initial peak density of $3 \times 10^{11}$ cm$^{-3}$, 4 $\mu$s of avalanche, expansion and dissociation to neutral atoms reduce the peak density to $4 \times 10^{10}$ cm$^{-3}$.  The average ion/Rydberg density in this ellipsoidal volume is $1.4 \times 10^{10}~{\rm cm}^{-3}$, and about 1.5 percent of the molecules in the plasma occupy two-thirds of its volume at a much lower density less than $10^{9}~{\rm cm}^{-3}$.

This Rydberg gas undergoes an avalanche to a quasi-neutral plasma of NO$^+$ and electrons at rate that rises sigmoidally in a time interval that varies depending on local density, from nanoseconds in the core of a higher-density ellipsoid to many microseconds in the periphery of an ellipse with a lower peak density \cite{review}.  

\subsubsection{Selective field ionization}

An electrostatic field ramp with a rise time of $\sim 0.8$ V cm$^{-1}$ ns$^{-1}$, started immediately after $\omega_2$ ($\Delta t_{ramp}=0$), drives a diabatic evolution of molecules in the $nf(2)$ Rydberg gas through the Stark manifold to cross a saddle point leading to ions and free electrons when the field $F$ in atomic units exceeds a threshold amplitude of ${1}/{9n^4}$.  In conventional units, this process forms a selective field ionization resonance beginning in V/cm at a field, $F =  (E_n(2)/4.59)^2$, where $E_n(2)$ in cm$^{-1}$ is the binding energy of the $nf$ Rydberg state with respect to a nitric oxide cation in rotational state, $N^+=2$.  

This trajectory through the Stark manifold traverses numerous intersections with states of matching electronic and rotational parity built on the ground rotational state of the ion.  By virtue of these crossings, the wavepacket acquires sufficient $N^+=0$ character to form free electrons and rotational ground state NO$^+$ cations earlier in the ramp, when the rising field passes an amplitude of $(E_n(0)/4.59)^2$ V/cm.

A ramp delayed by a few hundred nanoseconds samples the quantum-state distribution in an evolving Rydberg gas.  During the interval of this delay, promptly formed electrons collide with Rydberg molecules.  This causes $\ell$-mixing.  Molecules prepared in the initial state, $n_0f(2)$ change orbital angular momentum, populating a degenerate manifold of states, $\ket{N^+=2} \ket{n,\ell}$.  These states of higher orbital angular momentum field ionize at slightly higher field amplitudes to produce electron waveforms reflecting the formation of ions in rotational states, $N^+=0$ and 2.  

The integrated electron signal collected at a given ramp delay changes with the initial density of the Rydberg gas, $\rho_0$.  Normally, we saturate the $\omega_2$ transition and use the available density of NO A $^2\Sigma^+$ to regulate  $\rho_0$.  As described above and diagrammed in Figure \ref{fig:experiment}, we systematically control the density of the A-state molecules present for $\omega_2$ excitation by varying the $\omega_1$ pulse energy and, for a fixed pulse energy, by varying the  $\omega_1 - \omega_2$ delay.  Using these tools, we have confirmed over the range of the present experiment, that the integrated electron signal in an SFI trace depends in direct proportion on the initial density of the Rydberg gas.  Relying on this relationship, we systematically vary the A-state density and sort the many traces in a typical SFI contour at fixed ramp delay according to $\rho_0$.  

\begin{figure}
    \centering
    \includegraphics[width= .45  \textwidth]{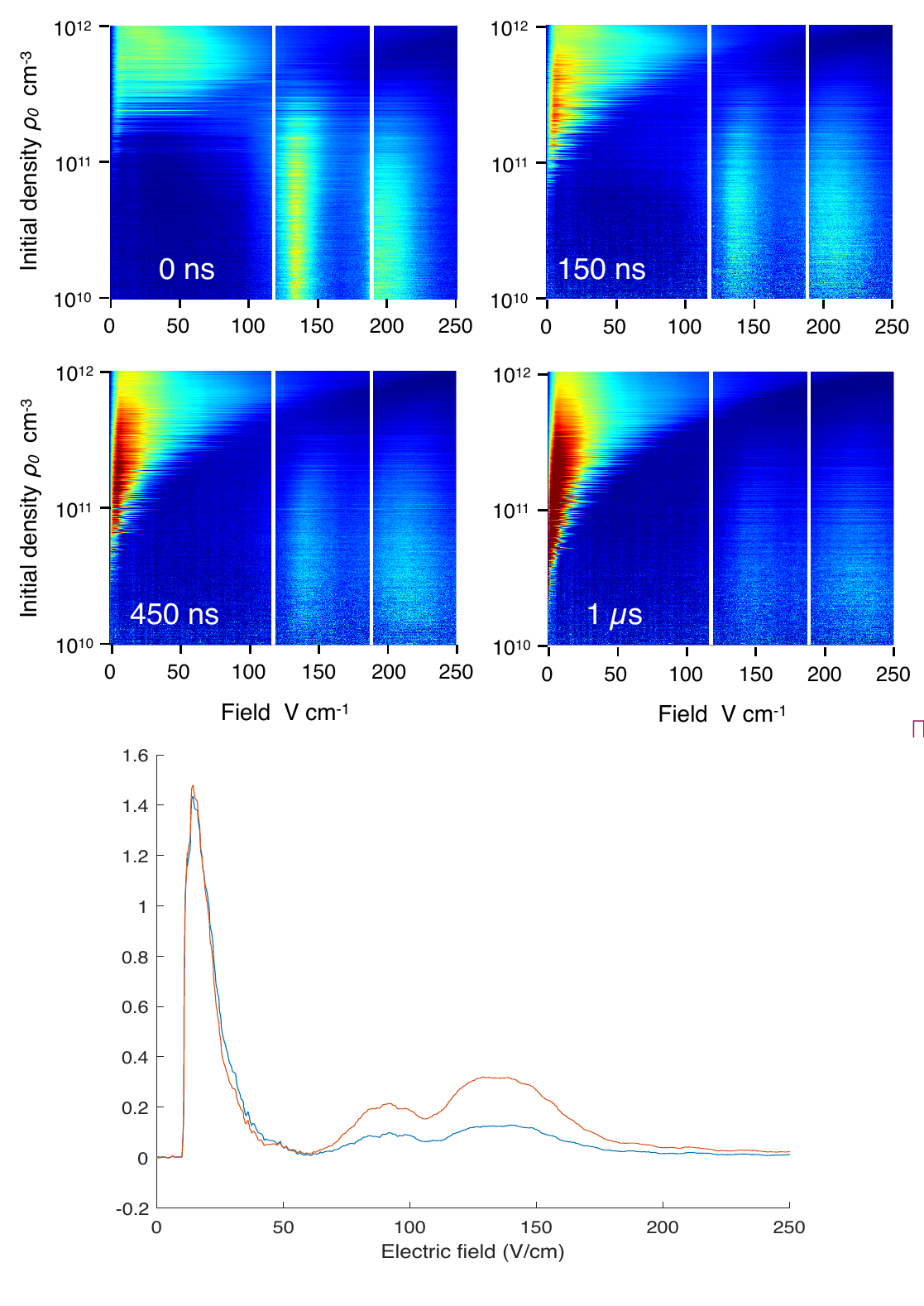}
    \caption{(upper frames) Typical SFI spectra, formed by  4,000 SFI traces sorted according to the initial density $\rho_0$, for an $nf(2)$ Rydberg with an initial principal quantum number $n_0 = 44$, following ramp delays, $\Delta t_{\rm ramp}$, of 0,150, 450 and 1000 ns. (bottom frame) Single SFI traces obtained under identical conditions of initial $49f(2)$ Rydberg gas density (10$^{11}$ cm$^{-3}$) and ramp-field delay (2 $\mu$s) in the absence (upper) and presence (lower) of a low-amplitude CW 60 MHz radio frequency field (V$_{\rm pp}=0.125$ V cm$^{-1}$).}
    \label{fig:SFI_3D}
\end{figure}

Selective field ionization spectra such as these provide a measure of the global spectrum of electron binding energy as a function of time.  Figure \ref{fig:SFI_3D} combines 4,000 SFI traces sorted by peak density over a range from 10$^{12}$ to 10$^{10}$ cm$^{-3}$.  The signal near zero field represents very high Rydberg molecules and electrons loosely-bound by the plasma space charge,  The two features that appear at higher field reflect the field ionization of the $n_0 = 44$ state to NO$^+~X~^1\Sigma^+$ cation rotational states, $N^+=0$ and $N^+=2$.  Note how after ramp delay of zero ($\Delta t_{\rm ramp} = 0$), these features shift to higher field ionization thresholds, reflective of electron collisional $\ell$-mixing of initial $44f(2)$ Rydberg molecules. 

Here, we see that that Rydberg gases with initial density in the higher range of our experiment avalanche fully on a 100 ns timescale, while Rydberg gases of lower density evolve to form a mixture of weakly bound NO$^+$ ions and electrons (ultracold plasma) together with Rydberg molecules that retain the initially selected principal quantum number, $n_0$.  Note the absence of electron signal in a wedge of very low SFI potential on the left-hand edge of these contours.  At the highest density, field ionization requires a minimum of a few V cm$^{-1}$, which corresponds to Coulomb binding energy on the order of 300 GHz.  

\begin{figure}
    \centering
     \includegraphics[width= .45  \textwidth]{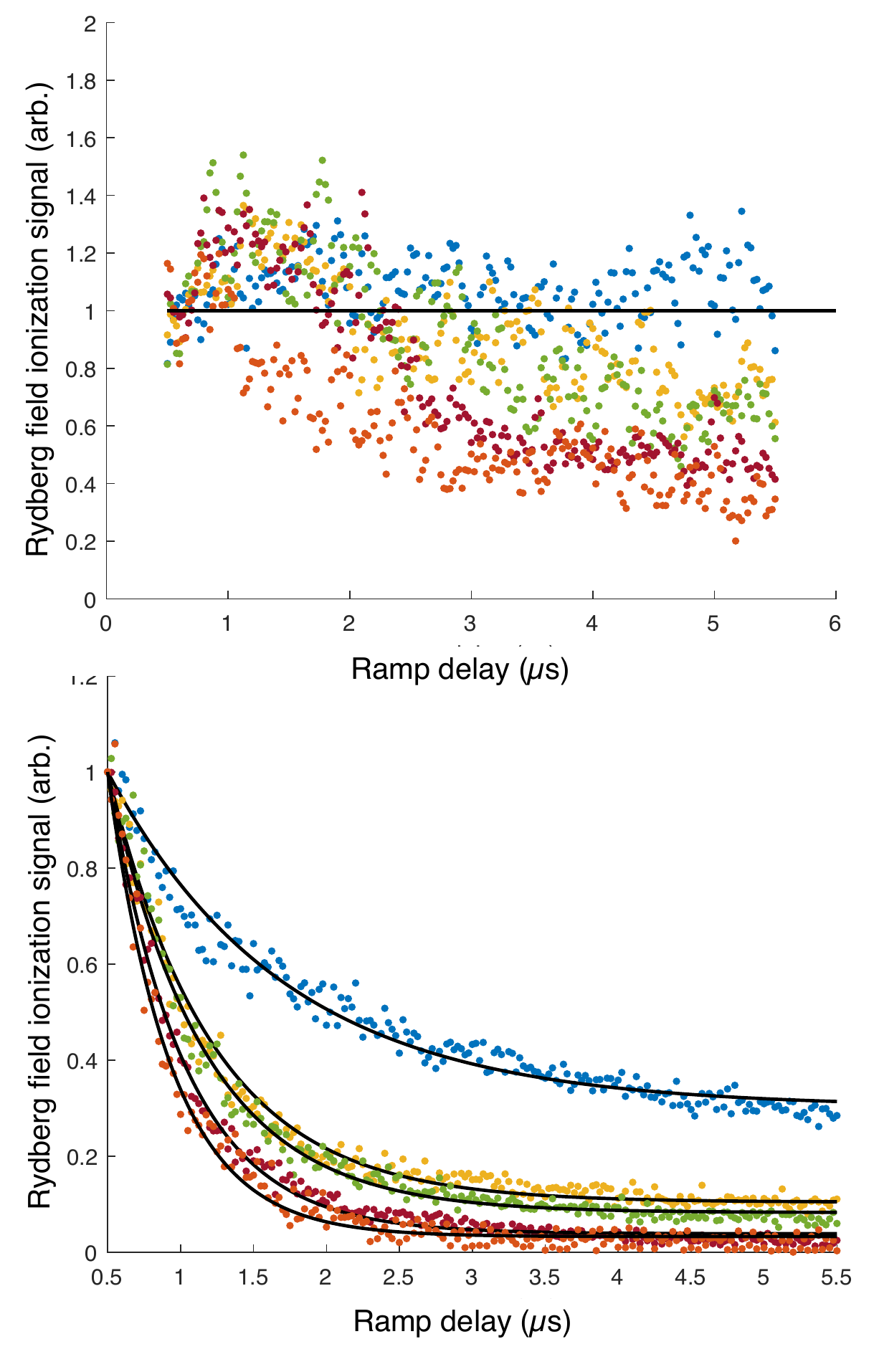}
    \caption{For ultracold plasmas evolving from $49nf(2)$ Rydberg gases:  (top) SFI amplitude integrated over ramp field from 0 to 50 V cm$^{-1}$ (plasma signal); and (bottom) SFI amplitude integrated over ramp field from 50 to 200 V cm$^{-1}$ (Rydberg signal), as a function of ramp delay, $\Delta t_{\rm ramp}$ under field-free conditions (blue) and in the presence of a 60 MHz rf pulse that starts 500 ns after $\omega_2$ excitation and extends to fill the elapsed time, $\Delta t_{\rm ramp}$, until the start of ramp field.  Peak-to-peak pulse amplitudes vary with increasing effect over the range, V$_{\rm pp}= 0.125$ (yellow), 0.25 (green), 0.5 (red), and 1.25 (orange) V cm$^{-1}$. }
    \label{fig:lifetimes}
\end{figure}

On a timescale of a few microseconds, the weakly bound ion-electron plasma population evolves differently from that of the $n_0$ Rydberg molecules.  Figure \ref{fig:lifetimes} compares the magnitude of the plasma signal, integrated over SFI ramp-field amplitudes from 0 to 50 V cm$^{-1}$, with that of the residual Rydberg population, observed for a higher range of ramp-field amplitudes from 50 to 200 V cm$^{-1}$.  

Note under field-free conditions that the number of electrons weakly bound to NO$^+$ cations and extracted by the leading edge of the SFI field ramp remains constant as we step the start of the ramp over a time from 500 ns to 5.5 $\mu$s after the $\omega_2$ excitation pulse that forms the Rydberg gas.  During this same time interval, the Rydberg signal falls exponentially, in each case to an apparent plateau.  

\subsection{Molecular ultracold plasma rate processes in the presence of a continuous wave (CW) radio frequency field}

Figure \ref{fig:lifetimes} also includes equivalent measurements made in the presence of a CW radio frequency field with a frequency of 60 MHz and varying peak to peak amplitude.  Here we see that the presence of an rf field of this frequency diminishes the plasma signal to a degree that depends on the peak-to-peak amplitude, V$_{\rm pp}$.  The rf field has a more dramatic effect on the decay of the $n_0$ Rydberg signal.  Over the range of amplitudes considered, the decay rate increases by a factor of from two to three and the plateau, which defines the fraction of $n_0$ Rydberg molecules that survive for very long times, decreases by a factor of ten.

This distinct effect of a radio frequency field on the SFI signal appears clearly in a single raw ramped field-ionization trace.  Figure \ref{fig:SFI_3D} shows the electron waveform obtained at intermediate density by the application of a field-ionization ramp delayed by 2 $\mu$s after $\omega_2$.  As in the accompanying contours, the low-voltage part of the ramp collects loosely-bound electrons from the plasma.  At higher voltage, we see the field ionization of the residual $n_0 = 44$ Rydberg gas to form NO$^+~X~^1\Sigma^+$ cation rotational states, $N^+=0$ and $N^+=2$.  Here, we also see the electron signal produced by the same ramp in the presence of a 60 MHz radio frequency field with an amplitude of 0.125 V cm$^{-1}$.  Note as above, on this timescale, the rf field has little effect on the weakly bound plasma waveform, but it substantially reduces the amplitude of the Rydberg signal.  

\subsection{Effects of a pulsed radio frequency field}

For a nominally constant initial Rydberg gas density, set by a fixed $\omega_1 - \omega_2$  delay of 200 ns, we find that a 250 ns pulsed 60 MHz radio frequency field applied to G$_2$ affects the $n_0$ Rydberg field ionization yield to a degree that varies with $\Delta t_{\omega_{\rm rf}}$, varied between zero (the time of $\omega_2$) and $\Delta t_{\rm ramp}$.  

\begin{figure}
    \centering
        \includegraphics[width= .45 \textwidth]{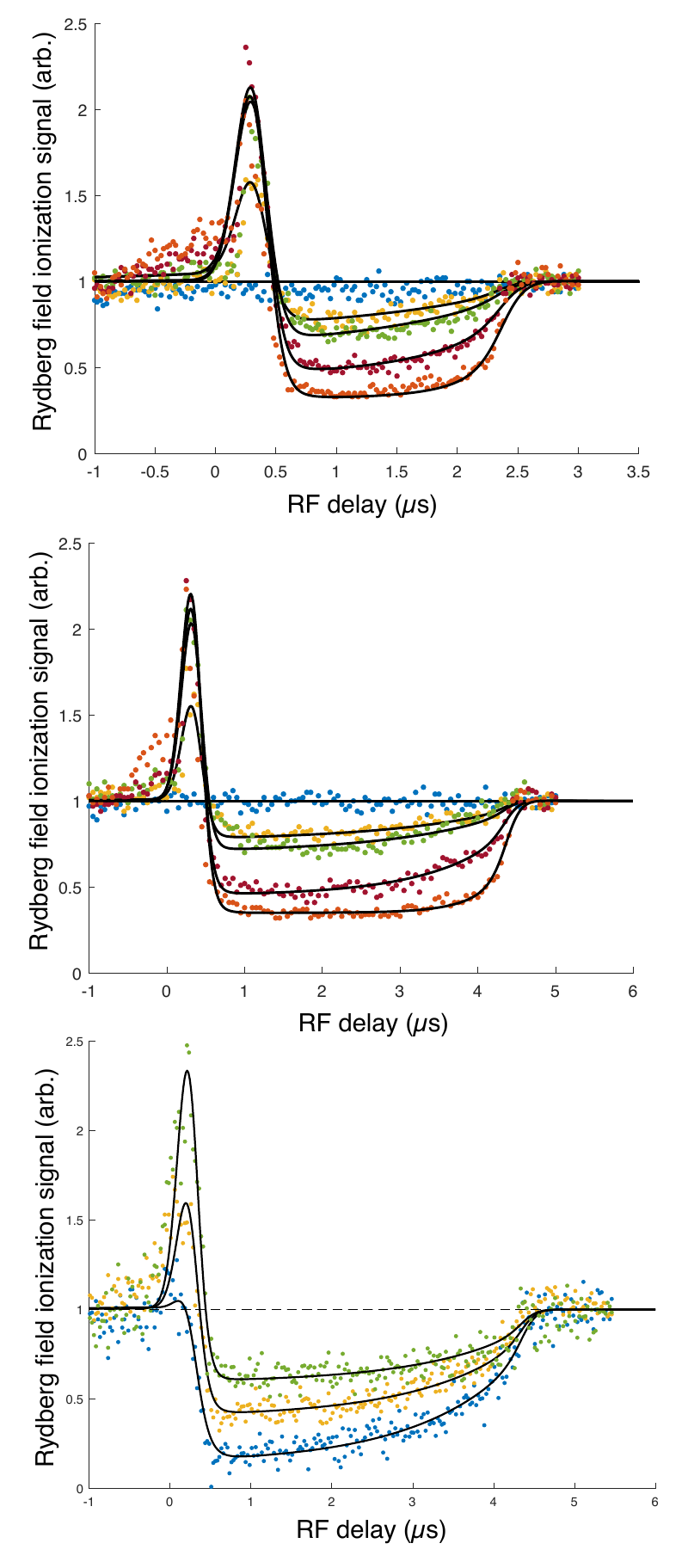}
     \caption{Integrated electron signal obtained by selective field ionization of an ultracold plasma as it evolves from a $49f(2)$ Rydberg gas of NO in the presence of a 250 ns  60 MHz pulsed radio frequency field, plotted as a function of rf delay, $\Delta t_{\omega_{\rm rf}}$   for five rf amplitudes, $V_{pp}$ in V cm$^{-1}$, 0 (blue), 0.125 (yellow), 0.25 (green), 0.5 (red) and 1.25 (orange), for a 200 ns $\omega_1 - \omega_2$ delay and two fixed values of ramp field delay, $\Delta t_{\rm ramp}$, of 2.8 $\mu$s  (top) and 4.8 $\mu$s (middle) after $\omega_2$ excitation.  (bottom) Fixed ramp field delay and $V_{pp}$ with three values of $\omega_1-\omega_2$ delay, 100 (blue), 200 (yellow) and 300 (green) ns. Each curve is normalized according to the signal at -1$\mu$s rf pulse delay when the rf pulse has no effect. }
    \label{fig:rf-delay}
\end{figure}

The SFI signal increases for an rf pulse applied during or just after $\omega_2$.  Beyond a delay of 500 ns, the application of a pulsed radio frequency field depletes the NO Rydberg molecule signal to a degree that depends on the temporal relation of the rf pulse to $\omega_2$ and the time at which the SFI ramp field on G$_1$ begins to rise.  

Figure \ref{fig:rf-delay} shows this effect for two different intervals of $\Delta t_{\rm ramp}$, 2.8 and 4.8 $\mu$s.   Here we see that the height of the signal increase for $\Delta t_{\omega_{\rm rf}} < 500$ ns and the depth of depletion for $\Delta t_{\omega_{\rm rf}} > 500$ ns both grow with increasing rf field amplitude, $V_{pp}$, from 0.125 to 1.25 V cm$^{-1}$.  

As the delay time of the radio-frequency pulse, $\Delta t_{\omega_{\rm rf}}$ nears the ramp delay, $\Delta t_{\rm ramp}$, the Rydberg signal recovers.  The steepness with which the Rydberg field ionization signal grows as $\Delta t_{\omega_{\rm rf}}$ approaches $\Delta t_{\rm ramp}$ measures the rate of rf-accelerated predissociation.  The rf-field accelerated component of NO Rydberg predissociation clearly increases with $V_{pp}$.  The degree of acceleration as a function of $V_{pp}$ matches for SFI ramp delays, $\Delta t_{\rm ramp} = 4.8$ and 2.8 $\mu$s.

The depth of the $n_0$ Rydberg SFI signal depletion for $\Delta t_{\omega_{\rm rf}} > 500$ ns, as well as the shape of its recovery as $\Delta t_{\omega_{\rm rf}}$ approaches $\Delta t_{\rm ramp}$, also vary with the initial density of the Rydberg gas.  Figure \ref{fig:rf-delay} shows examples obtained for $\omega_1 - \omega_2$  delays of 100, 200 and 400 ns.  Here, for a constant rf amplitude, $V_{pp}=0.4$ V cm$^{-1}$, we see the least depletion at the lowest initial Rydberg gas density (obtained for the longest $\omega_1-\omega_2$ delay).  Less obviously, the apparent predissociation rate measured by the rise in the signal as $\Delta t_{\omega_{\rm rf}}$ approaches $\Delta t_{\rm ramp}$ lessens in an ultracold plasma of lower density.

Note, as seen for the experiment varying $V_{pp}$, the $n_0$ Rydberg SFI signal increases in the presence of an rf field for $\Delta t_{\omega_{\rm rf}} < 500$ ns.  Here, for constant $V_{pp}=0.4$ V cm$^{-1}$, an early pulsed rf field produces the greatest proportional enhancement by far for the lowest initial Rydberg gas density.  When applied in the first 250 ns after $\omega_2$, a radio frequency field with an amplitude of 400 mV cm$^{-1}$ causes very little if any enhancement in the SFI signal obtained for the Rydberg gas with a comparatively high initial-density, formed by using an $\omega_1-\omega_2$ delay of 100 ns.

\section{Discussion}

\subsection{Field-free avalanche dynamics and dissociation in the state-selected nitric oxide Rydberg gas}

Double-resonant excitation of nitric oxide at moderate density in a seeded supersonic molecular beam entrains an ellipsoidal volume of Rydberg gas that undergoes avalanche to form an ultracold plasma.  Electron signal contours in sorted selective field ionization spectra such as those evident in Figure \ref{fig:SFI_3D} show how the electron binding dynamics of this avalanche vary with the initial density of the Rydberg gas.  Here we see snapshots of the ultracold plasma electron binding energy spectrum formed by $44f(2)$ Rydberg gases ranging over two orders of magnitude in initial density after evolution times selected by $\Delta t_{\rm ramp}$ values of 0, 150, 450 and 1000 ns.  

For initial densities exceeding $ 5 \times10^{11}$ cm$^{-3}$, virtually all of the charge extracted by an applied ramp appears at relatively low field, over a range that extends from about 5 to no more than 80 V cm$^{-1}$.  The collection of these easily extracted electrons at zero ramp-delay signals a prompt avalanche that promotes the entire system to a state composed of high-Rydberg molecules and/or quasi-free electrons bound by the space charge of NO$^+$ ions.  

A Rydberg gas of moderate density, on the order of $3 \times 10^{11}$ cm$^{-3}$, also avalanches to yield electrons of low binding energy, mixed here with a residual population of Rydberg molecules bound with an energy determined by the initial principal quantum number, $n_0$.  This signature of $n_0$ Rydberg molecules, which field-ionize to form NO$^+$ in N$^+ = 0$ and 2, dominates the SFI spectrum of lower-initial-density Rydberg gases when measured with short ramp delays.  For all densities, the particle balance in this ultracold plasma shifts after a longer evolution time to favour electrons weakly bound to NO$^+$ ions, as high-$n$ Rydberg molecules or as electrons bound to the plasma space charge.  

Inspection of SFI spectra between 0 and 150 ns shows direct evidence of initial electron mobility in a subtle shift of residual $n_0$ features to higher appearance potential.  This shift to higher field-ionization threshold points to a dynamic process in which electron-Rydberg collisions drive $\ell$-mixing interactions that cause the photo-selected $44f(2)$ molecules to spread over the full distribution of $\ell \in {(0,1,2 ... 43)}$ values of Rydberg orbital angular momentum for $n_0=44$ \cite{WalzFlannigan}.  

From previous work \cite{Haenel2017}, we know that the overall density of the nitric oxide molecular ultracold plasma falls with evolution time during the first few microseconds, owing to channels of neutral decay via NO$^+$ ion-electron dissociative recombination and neutral NO Rydberg predissociation \cite{PCCP,Saquet2012,Sadeghi:2014}.   Figure \ref{fig:lifetimes} details this effect experimentally.  Note that the residual $n_0$ Rydberg signal decays, while the plasma signal remains constant on a microsecond timescale.  This directly shows that $n_0$-Rydberg predissociation dominates ultracold plasma decay on the timescale of this observation.   

However, after a period of time that varies to some degree with Rydberg gas density and initial principal quantum number, the $n_0$ Rydberg molecules sampled by SFI cease to predissociate.  Figure \ref{fig:SFI_3D} shows this effect distinctly in the residual $n_0$-Rydberg signal evident after a delay of 1 $\mu$s in the SFI spectrum of a $44f(2)$ Rydberg gas with an initial density of 10$^{11}$ cm$^{-3}$.  This appears more evidently in the exponential decay of the integrated $n_0 = 49$ Rydberg signal as a function of ramp field delay in Figure \ref{fig:lifetimes}, which in the absence of a radio frequency field falls to a residual plateau.  

Recognizing this, we fit the data in Figure \ref{fig:lifetimes} to a rate law of the form:
\begin{equation}
\frac{d\left [ {\rm NO^*} \right ]}{dt} = k_{\rm PD}\left [ {\rm NO^*} \right ] + \left [ {\rm NO^*} \right ]_A
\label{eqn:ratelaw}
\end{equation}
where $\left [ {\rm NO^*} \right ]$ describes the density of predissociating $n_0$ Rydberg molecules, $\left [ {\rm NO^*} \right ]_A$ represents the residual population in a state of arrested predissociation.  Here, $k_{\rm PD}$ refers to a phenomenological overall rate of predissociation.  Table \ref{rampdelay} gives these parameters for the fits plotted in Figure \ref{fig:lifetimes}.   

\begin{table*}
\caption{Kinetic parameters used in Eq (\ref{eqn:ratelaw}) to fit the exponential decay in the $n_0 = 49$ Rydberg molecule SFI signal in Figure \ref{fig:lifetimes}  under  field-free conditions and in the presence of a 60 MHz radio frequency field of varying peak-to-peak amplitude, $V_{pp}$. \\} 
\label{rampdelay}
\centering
\begin{tabular}{cccc}
\toprule
 $V_{pp}$ (V cm$^{-1}$ ) & \hspace{20 pt} $\left [ {\rm NO^*} \right ]_0$  & \hspace{20pt}  $k_{\rm PD}$ ($\mu$s$^{-1}$) & \hspace{8pt}  $\left [ {\rm NO^*} \right ]_A / \left [ {\rm NO^*} \right ] _0 $ \\
 \hline
  \multicolumn{4}{c}{Field Free}\\
 0.000  & \hspace{20 pt} 1 & \hspace{20 pt} 0.82  \hspace{8pt}  & 0.30 \\
   \multicolumn{4}{c}{60 MHz}\\
 0.125  & \hspace{20 pt}  1 & \hspace{20 pt} 1.39   \hspace{8pt}  & 0.10 \\
0.250   & \hspace{20 pt}  1 & \hspace{20 pt} 1.51   \hspace{8pt}  & 0.08 \\
0.500   & \hspace{20 pt}  1 & \hspace{20 pt} 1.90   \hspace{8pt}  & 0.04 \\
1.250   & \hspace{20 pt}  1 & \hspace{20 pt} 2.30   \hspace{8pt}  & 0.03 \\
\hline
 \end{tabular}
\end{table*}

%Thus the presence of residual long-lived Rydberg molecules at the initially selected principal quantum number would suggest the absence of $\ell$-changing electron-Rydberg collisions in the long-time dynamics of this plasma.  

%But, without $\ell$-mixing, higher-$\ell$ states survive to serve as a probe of any induced collisional redistribution.  

\subsection{Nitric oxide Rydberg predissociation in a regime of $\ell$-mixing}

Previous studies by Vrakking and Lee have established that a nitric oxide Rydberg molecule in the $n_0f(2)$ series near $n_0=49$ predissociates with a field-free lifetime of 10 ns \cite{Vrakking1995pra,Vrakking1995jcp}.  A dc field of a few hundred mV mixes $\ell$ sufficiently to increase this lifetime to a measured 75 ns.  Multichannel effective Hamiltonian models predict that effective $k_{\rm PD}$ values for fully coupled bright states in this range of the $nf$ series decrease to $\sim 5$ $\mu{\rm s}^{-1}$ \cite{Bixon,Remacle1998}.  Predissociation in the broader manifold of $n_0=49$ Rydberg states, mixed by electron collisions over all values of $\ell$, proceeds with a phenomenological $k_{\rm PD}$ determined by sampling $\ell$-detailed rates.  

 Referring to work cited above, we can assume that the Rydberg states of nitric oxide predissociate with characteristic rate constants, $k_{\rm PD}$, that fall systematically with increasing $n$ as $1/n^3$.  For a given $n$, $k_{\rm PD}$ depends very sensitively on orbital angular momentum $\ell$.  Only low-$\ell$ states decay with appreciable rates.  For the purposes of illustration, we can take rates from a model developed by Gallagher and coworkers \cite{Murgu2001}, patterned on the work of Bixon and Jortner \cite{Bixon}. and estimate $k_{\rm PD}$ for a given $n$ from the statistically weighted sum of $\ell$-dependent rate constants, $k_\ell=0.014$, 0.046, 0.029 and 0.0012 in atomic units for $\ell$ from 0 to 3, and 0.00003 for $\ell \geq4$, scaled by $n^{-3}$: 
\begin{equation}
k_{\rm PD}(n) = \frac{\sum_\ell{(2\ell+1)k_{n,\ell}}}{n^2}\frac{4.13\times 10^{16} {\rm s^{-1}}}{2\pi n^3}
\label{kPD}
\end{equation}
\noindent This simple statistical approach predicts decay times for levels near $n=50$ of about 200 ns, in accord with observations for bright states of NO in this range, when prepared by broad-band excitation in the presence of an $\ell$-mixing electric field \cite{Vrakking1995pra,Bixon,Remacle1998}.  

The initial predissociation kinetics observed in the present experiment conform with this picture.   Coupled differential equations describing inelastic electron-Rydberg collisional evolution in principal quantum number, $n$, and electron-impact ionization, together with three body electron-ion recombination accurately account for the the first 500 ns in the field-free relaxation of a nitric oxide Rydberg gas to plasma as a function of initial density, $\rho_0$, and initial principal quantum number, $n_0$ \cite{Saquet2012,Haenel.2018}.  For the particular initial density represented in Figure \ref{fig:lifetimes}, we find that $n_0$ nitric oxide molecules in this ultracold plasma initially decay to neutral products on a timescale consistent with the state-detailed rate of NO Rydberg predissociation in a collisional regime of $\ell$ scrambling. 

However, as evident in Figure  \ref{fig:lifetimes} and the fits to Eq (\ref{eqn:ratelaw}) parameterized in Table \ref{rampdelay}, the molecular nitric oxide ultracold plasma displays a persistent residual population of $n_0$ Rydberg molecules that survives the avalanche of a state-selected $n_0$ Rydberg gas to plasma and the quench of this plasma to a state of very low electron binding energy. 

We can explain this apparently cold, arrested state as evidence for the presence of a long-lived high-$\ell$ residue of $n_0$ Rydberg molecules that remains from the statistical distribution over all accessible values of $\ell$ created by electron-collisional $\ell$-mixing during the avalanche.  Such a residue can survive only if $\ell$-mixing ceases.  Its presence here serves as an adventitious sensor of $\ell$-mixing under these conditions of arrested relaxation.  

This enduring population of long-lived $n_0$ Rydberg molecules thus suggests that the system evolves to a state of quenched predissociation in which the ultracold plasma contains too few free electrons to $\ell$-mix these residual $n_0$ Rydberg molecules.  We might explain this by assigning all the signal that appears in the prominent contours at low field in the SFI spectrum to electrons bound in very high-$n$ Rydberg states.  

However, for a initial principal quantum in the range of $n_0 \approx 44$, classical rate theory considerations call for strong Rydberg-Rydberg interactions, Penning ionization and electron-ion-Rydberg molecule collisions, giving rise to an $\ell$-mixed quasi-equilibrium within 100 $\mu$s at a density as low as 10$^8$ cm$^{-3}$ \cite{WalzFlannigan}.  Coupled rate-equation simulations predict that such interactions drive Penning ionization and avalanche to plasma in Rydberg gas systems with the particular range of densities displayed in Figure \ref{fig:SFI_3D} on a microsecond timescale \cite{Haenel.2018}.

\subsection{Effect of a radio-frequency field} \label{field-effect}

The nitric oxide molecular ultracold plasma evolves in the long-time limit to a state in which most of the electrons bind very weakly to ions, either individually in very high-$n$ Rydberg orbitals or collectively to the NO$^+$ space charge \cite{Haenel2017}.  Either way, the SFI experiment directly measures an ultracold plasma electron binding energy no greater than $\sim 800$ GHz.  A small portion of this system survives as Rydberg molecules with the initially selected principal quantum number, $n_0$, in a high-$\ell$ state of quenched predissociation, suggesting an absence of $\ell$-mixing electron-Rydberg collisions.   

The application of a CW 60 MHz field with a peak-to-peak amplitude in the range from 0.125 to 1.25 V cm$^{-1}$ evidently accelerates the predissociation of these residual $n_0$ Rydberg molecules.  A radio-frequency field interacts with a conventional plasma of electrons and ions to drive collective modes of motion termed plasma oscillations.  A neutral plasma of defined density, $\rho_e$, supports a plasma frequency $\omega_{\rm rf} = \sqrt{e^2\rho_e/\epsilon_0 m_e}$.  A field with a frequency of 60 MHz resonates with an electron-ion plasma at a density of about 10$^6$ cm$^{-3}$.  The density of the nitric oxide molecular ultracold plasma volumes sampled by selective field ionization falls in a calibrated range from 10$^{10}$ to 10$^{12}$ cm$^{-3}$.  The observed conditions of plasmon resonance therefore classify the carrier of the measured SFI signal as a dielectric phase suspended in a rarified conducting background.     

A radio frequency field of an amplitude in the range from 0.125 to 1.25 V cm$^{-1}$ drives plasma oscillations in this conducting background.  This appears to cause a resumption of $\ell$-mixing that redistributes Rydberg orbital angular momentum from states of high-$\ell$ to low-$\ell$ states of shorter predissociation lifetime.  As detailed above, it is quite reasonable to attribute this rf-driven $\ell$-mixing to renewed electron transport in the bulk and an avalanche-like effectiveness of electron collisions \cite{WalzFlannigan}.

\subsection{Kinetics of radio-frequency accelerated predissociation} 

Nitric oxide Rydberg molecules with the originally selected principal quantum number, $n_0$, predissociate with detailed unimolecular rate constants $k_{n_0,\ell}$.  In a statistical limit, these rate constants combine in accordance with Eq (\ref{kPD}) to determine a phenomenological rate constant, $k_{\rm PD}(n_0)$ for the residual population of $n_0$ Rydberg molecules.  

Under field-free conditions, Figure \ref{fig:lifetimes} shows that the observed predissociation rate constant varies in time, and ultimately falls to zero, leaving an arrested population of $n_0$ Rydberg molecules, $\left [ {\rm NO^*} \right ]_A$.  This is easily explained.  Without efficient redistribution in $\ell$, $k_{\rm PD}(n_0)$ deviates from the value predicted by  Eq (\ref{kPD}) for a statistical distribution of Rydberg angular momentum.  As Rydberg molecules of lower-$\ell$ predissociate, the distribution of population over $\ell$ shifts.  Coefficients of the largest terms in the sum of detailed rate constants that determines $k_{\rm PD}(n_0)$ fall to zero, and predissociation ceases.  

The presence of a CW radio-frequency field appears to sustain $\ell$-mixing.   Phenomenologically the presence of the field serves to maintain the weight of low-$\ell$ terms in Eq (\ref{kPD}), sustaining a low-$\ell$ contribution to the phenomenological rate constant $\delta k_{\rm PD}(n_0)$ and diminishing the fraction of arrested molecules.  Table \ref{rampdelay} quantifies these trends for the conditions of the experiment represented in Figure \ref{fig:lifetimes}, where a contribution to the predissociation rate process owing to low-$\ell$ predissociation doubles or triples the phenomenological rate constant, $k_{\rm PD}(n_0)$, and reduces the arrested fraction, $\left [ {\rm NO^*} \right ]_A / \left [ {\rm NO^*} \right ] _0 $, to nearly zero, depending on $V_{pp}$.  

A pulsed radio frequency field similarly promotes a redistribution of residual $n_0$ Rydberg molecules over $\ell$, with a comparable effect on the apparent rate of predissociation.  We see this in a decreased $n_0$ Rydberg contribution to the SFI spectrum.  Let us assume that a radio frequency pulse applied anytime after avalanche and electron-collisional $\ell$-mixing produces the same extent of redistribution over $\ell$, and as a result, increases $k_{\rm PD}(n_0)$ to the same degree.  Accelerated predissociation then acts to diminish the $n_0$ Rydberg signal by an amount that depends on the elapsed time between the start of the radio frequency pulse, $\Delta t_{\omega_{\rm rf}}$, and the beginning of the SFI ramp field at $\Delta t_{\rm ramp}$.  

An advancing $\Delta t_{\omega_{\rm rf}}$, and thus smaller gap, $\Delta t_{\rm ramp} - \Delta t_{\omega_{\rm rf}}$, gives an rf-accelerated predissociation rate constant less time to act and thus causes a smaller suppression of the $n_0$ Rydberg signal.  We represent this effect by developing a phenomenological expression for the density of Rydberg molecules as a function of $\Delta t_{\omega_{\rm rf}}$ normalized by its value at a time, $t = \Delta t_{\rm ramp}$ under field-free conditions for any initial density, as determined say by the $\omega_1 - \omega_2$ delay.  

In a limit of instantaneous rf-induced $\ell$-mixing and ultrafast NO Rydberg predissociation, the normalized Rydberg field ionization signal depleted by a pulsed radio frequency field would recover to a value of 1 as the convolution of the rising edge of the rf pulse and the SFI voltage ramp.  For present purposes, we can arbitrarily approximate this by a logistic function, with arbitrarily chosen onset and rise time parameters, $[t_r, k_r]$:
\begin{equation}
S_e = (1-f_e) + f_e \left [ \frac{1}{1+ e^{k_r(t_r - t)} }\right ],
\label{equ:step}
\end{equation}

Here, $f_e$ represents the full fractional extent to which a radio frequency pulse of peak-to-peak amplitude, $V_{pp}$ depletes the $n_0$-Rydberg SFI signal measured by a voltage ramp that starts at  $\Delta t_{\rm ramp}$.  If the predissociation proceeds at a rate slower than instantaneous, then a pulsed radio-frequency field applied late in time, $\Delta t_{\omega_{\rm rf}}$, close to $\Delta t_{\rm ramp}$, will act with less effectiveness in depleting the $n_0$-Rydberg signal than an rf field applied earlier.  

We can account for less-than-instantaneous predissociation if we regulate the depletion of $S_e$ to a degree determined by an amount added to the predissociation rate constant, $\delta k_{\rm PD}$ and the time difference between $t = \Delta t_{\omega_{\rm rf}}$ and $\Delta t_{\rm ramp}$.  
\begin{align}
&S_e = (1-f_e) + f_e \left [ \frac{1}{1+ e^{k_r(t_r - t)} }\right ] \nonumber \\
&+ f_e\left[ e^{-\delta k_{\rm PD}(\Delta t_{\rm ramp}-t)} \right] \times \left (1- \left [ \frac{1}{1+ e^{k_r(t_r - t)} }\right ]  \right )
\label{equ:fullstep}
\end{align}
where we moderate the effect of the exponentially recovered $n_0$ Rydberg signal by a step function that falls from 1 to 0.  Here, we assign a uniform midpoint, $t_0$, to the rising and falling logistic functions that best represent the convolution of the rf pulse with the ramp.  We then fit $f_e$ and $\delta k_{\rm PD}$ as they vary with density.  

Just as late $\ell$-mixing diminishes the $n_0$-Rydberg SFI signal for $\Delta t_{\omega_{\rm rf}} > 500$ ns by redistributing high-$\ell$ to low, early $\ell$-mixing at $\Delta t_{\omega_{\rm rf}} $ times less than 500 ns preserves $n_0$-Rydberg population by accelerating the promotion of $n_0f(2)$ states selected by double resonant laser excitation to $n_0$ orbitals of higher $\ell$.  Experiments some time ago in the Gallagher group spectroscopically quantized a similar effect in low-density Rydberg gases of NO in the present range of principal quantum number \cite{Murgu2001}.  

For the simple purpose of describing the observed degree of rf enhancement for $\Delta t_{\omega_{\rm rf}} < 500$ ns, we fit the data phenomenologically with an exponential rise and logistic decay:
\begin{equation}
S_0=\left( f_e+f_0 e^{-\delta k_0(t- t_0)^2} \right) 
 \times\left (1- \left [ \frac{1}{1+ e^{k_r(t_0 - t)} }\right ]  \right ) 
\label{equ:avalanche}
\end{equation}

Curves drawn through the data in Figure \ref{fig:rf-delay} sum $S_0$ contributions owing to $\ell$-mixing Rydberg lifetime enhancement during the avalanche, Eq (\ref{equ:avalanche}), together with $S_e$ contributions owing to $\ell$-mixing Rydberg lifetime depletion during the subsequent phase of arrested relaxation, Eq (\ref{equ:fullstep}), as a function of $\Delta t_{\omega_{\rm rf}}$ for fixed ramp field delays, $\Delta t_{\rm ramp}$, of 2.8 and 4.8 $\mu$s.  

We recognize that arbitrary logistic functions imperfectly describe convolutions of varying degrees of $\ell$-mixing and changes in $k_{PD}$ as the rf pulse passes through $\omega_2$ and the SFI ramp.  Nevertheless, we readily obtain self-consistent representations of the SFI signal as a function of $\Delta t_{\omega_{\rm rf}}$ at both ramp field delays, changing only the known fixed value of  $\Delta t_{\rm ramp}$.  Table \ref{rfdelay} summarizes the parameters of these fits.  Note that we succeed in describing the recovery of the signal as the rf pulse passes through the ramp by a logistic function with the same offset, $\Delta t_{\rm ramp} - t_r$ at both ramp-field delays.

\begin{table*}
\caption{Parameters used in Eqs (\ref{equ:fullstep}) and (\ref{equ:avalanche}) to describe the enhancement, suppression and recovery of the $n_0$ Rydberg SFI signal by the action of a 250 ns 60 MHz pulsed radio frequency field of peak-to-peak amplitude, $V_{pp}$, as time of this pulse, $\Delta t_{\omega_{\rm rf}}$ advances from $t_{\omega_2}$ to pass through the beginning of the SFI ramp field at $\Delta t_{\rm ramp}$ for ramp field delays of 2.8 and 4.8 $\mu$s. \\} 
\label{rfdelay}
\centering
\begin{tabular}{ccccccccccc}
\toprule
$V_{pp}$ & $\Delta t_{\omega_2}$  & $\rho_0$ & $k_r$ & $t_0$ & $t_r$  & $\Delta t_{\rm ramp} $ & \hspace{5 pt} {$f_0$} & \hspace{5 pt}$f_e$ & \ $\delta k_{0}$  & $\delta k_{\rm PD}$  \\ 
(V cm$^{-1}$)&(ns)&($\mu$m$^{-3}$)&($\mu$s$^{-1}$) &($\mu$s) &($\mu$s) &($\mu$s) &&&($\mu$s$^{-1}$) &($\mu$s$^{-1}$) \\
\hline
\multicolumn{11}{c}{2.8 $\mu s$ ramp field delay} \\
0.125 & 200 & 0.35 & 10 & 0.38 & 2.4 & 2.8 & 1.10 &  \hspace{10 pt}0.31 & 20 & 0.65 \\
0.250& 200 & 0.35 & 10 & 0.38 & 2.4 & 2.8 & 1.82 &  \hspace{10 pt}0.41 & 20 & 0.75 \\
0.500& 200 & 0.35 & 10 & 0.38 & 2.4 & 2.8 & 2.10 &  \hspace{10 pt}0.55 & 20 & 1.40 \\
1.250 & 200 & 0.35 & 10 & 0.38 & 2.4 & 2.8 & 2.10 &  \hspace{10 pt}0.68 & 20 & 2.58 \\ 
\hline
\multicolumn{11}{c}{4.8 $\mu s$ ramp field delay} \\
0.125 & 200 & 0.35 & 10 & 0.4 & 4.4 & 4.8 & 1.10 & 0.23 & 20 & 0.64 \\
0.250& 200 & 0.35 & 10 & 0.4 & 4.4 & 4.8 & 1.83 & 0.30 & 20 & 0.70 \\
0.500 & 200 & 0.35 & 10 & 0.4 & 4.4 & 4.8 & 2.10 & 0.55 & 20 & 1.01 \\
1.250 & 200 & 0.35 & 10 & 0.4 & 4.4 & 4.8 & 2.30 & 0.65 & 20 & 2.34 \\ 
 \hline
\multicolumn{11}{c}{4.8 $\mu s$ ramp field delay, varying initial density} \\
0.400 &100 & 0.59 & 10 & 0.3 & 4.4 & 4.8 & 0.33 & 0.86 & 20 & 0.8  \\
0.400 &200 & 0.35 & 10 & 0.3 & 4.4 & 4.8 & 1.24 & 0.60 & 20 & 0.8   \\
0.400 &300 & 0.21 & 10 & 0.3 & 4.4 & 4.8 & 2.38 & 0.41 & 20 & 0.8   \\ 
\bottomrule
 \end{tabular}
\end{table*}

Table \ref{rfdelay} quantitatively represents the clear trends displayed in Figure \ref{fig:rf-delay}.  A radio frequency field of growing peak-to-peak amplitude increases the early-time amplitude of the $n_0$-Rydberg signal enhancement ($f_0$), increases the later-time amplitude of the $n_0$-Rydberg signal depletion ($f_e$), and increases the degree to which the rate of predissociation is enhanced ($\delta k_{PD}$)

\subsection{Mechanics of radio frequency induced $\ell$-mixing} 

The nitric oxide ultracold plasma evolves to a state of suppressed Rydberg predissociation, marked here by the persistence of a residual population of high-$\ell$ Rydberg molecules that retain the initially selected principal quantum number, $n_0$.  This occurs as the consequence of a process that begins in the avalanche with $\ell$-mixing electron-Rydberg collisions.  This randomization in $\ell$ is followed by evolution in the bulk plasma to a condition both of low electron binding energy and quenched electron mobility.  This quenched environment traps a measurable fraction of $n_0$ Rydberg molecules in states of $\ell$ too high to predissociate.  

We could interpret sequences of SFI spectra, such as those shown in Figure \ref{fig:SFI_3D} as evidence for the classical evolution of the NO Rydberg gas to an ultracold plasma background consisting entirely of very high-$n$ Rydberg molecules.  Assuming such states were stable, this very high-$n$ Rydberg gas background would contain no free electrons.  In the absence of Penning ionization and avalanche in this background, residual $n_0$ Rydberg molecules of low-$\ell$ would predissociate, and a very long-lived high-$\ell$ ensemble of Rydberg molecules would remain,  consistent with the observed field-free state of arrested predissociation.  
 
The evident perturbation of this arrested system by a 60 MHz radio frequency field would appear to require free electrons activated by plasma oscillations, and thus oppose this Rydberg gas scenario.  However, Gallagher and coworkers have shown that a radio frequency field alone can drive transitions that scramble the distribution over $\ell$ within a single Stark manifold \cite{Murgu2001}.  In their experiment on low-$\ell$ Rydberg states of NO, this effect lengthened predissociation lifetimes.  The same mechanism of $\ell$-mixing could accelerate predissociation in an arrested distribution of residual $n_0$ Rydberg molecules of high-$\ell$.  

Coupled rate-equation simulations suggest that Penning ionization and avalanche occur too quickly at our density for collision-free rf excitation to serve as the leading cause of $\ell$-mixing in the present case \cite{Haenel.2018}.  For greater certainty, we refer to an experimental result presented above that tells us directly whether the rf-depletion observed here arises from electron collisions or field-induced $\ell$-mixing.  

Predissociation stimulated by Stark mixing in a radio frequency field of a given amplitude occurs to the same degree for every molecule in a sample of any density.  Predissociation catalyzed by $\ell$-mixing electron collisions proceeds as a pseudo first-order process, and thus occurs to a fractional extent that varies with the density of electrons.  Figure \ref{fig:rf-delay} shows immediately that the fractional depletion changes with the initial density of the Rydberg gas, controlled with precision by adjusting the $\omega_1 - \omega_2$ delay ($\Delta t_{\omega_2}$).  As shown in Table \ref{rfdelay} curves through these data, obtained for densities that differ by about a factor of three, fit Eq (\ref{equ:fullstep}) varying only $f_e$.  

Note that the initial Rydberg gas density affects only the fractional depletion.  The additive contribution to $k_{\rm PD}$ does not vary over the range of charged particle densities formed as a consequence of the variation of initial Rydberg gas density.  This suggests that the scrambling in $\ell$ saturates after very few electron-Rydberg collisions.

It is instructive to consider the effect of an rf pulse applied for $\Delta t_{\omega_{\rm rf}} < 500$ ns on the $n_0$-Rydberg SFI signal under conditions of constant density and varying peak-to-peak amplitude, $V_{pp}$ (Figure \ref{fig:rf-delay}) compared with an rf pulse of constant $V_{pp}$ applied under conditions of varying Rydberg gas density (Figure \ref{fig:rf-delay}).  

In both cases, we can understand the effect of the radio frequency field as one of facilitating electron-Rydberg collisions.  At early times, when most Rydberg molecules still occupy the optically selected $n_0f(2)$ Rydberg state, electron collisions lead to lifetime lengthening, by scattering $\ell=3$ Rydberg molecules to states of higher $\ell$.  

Preceding conditions match for the systems of constant density (Figure \ref{fig:rf-delay}), and enhancement at early times occurs in an equivalent growing background of electrons released by Penning ionization and avalanche.  The radio frequency field of largest amplitude has greatest effect on the Rydberg population both before and after collisional $\ell$-mixing is complete.  The highest $V_{pp}$ causes the largest enhancement for $\Delta t_{\omega_{\rm rf}} < 500$ ns and deepest depletion for $\Delta t_{\omega_{\rm rf}} > 500$ ns,  The Table \ref{rfdelay} amplitudes, $f_0$ and $f_e$ increase together with increasing  $V_{pp}$.   

Under conditions of constant $V_{pp}$ but varying density, the initial Penning electron density and avalanche rate differ substantially, as can be read directly in Figure \ref{fig:rf-delay}.  The system of lowest density sustains a comparatively low avalanche electron density and so sees a comparatively smaller degree of rf-induced depletion.  But, its early lingering population of $n_0f(2)$ Rydberg molecules exhibits a large degree of rf-accelerated $\ell$-mixing and stabilization.  

The opposite situation applies in the system of highest density.  With a large avalanche electron density, the second-order amplitude of rf-induced $\ell$-mixing and predissociative SFI signal loss is great.  But with a rapid avalanche and nearly complete degree of early electron-collisional $\ell$-mixing, the system shows very little SFI signal increase as the rf pulse scans through $\Delta t_{\omega_{\rm rf}} < 500$ ns.  In Table \ref{rfdelay}, the amplitudes, $f_0$ and $f_e$ vary in opposite directions with density.

\subsection{Electron mobility in the quenched molecular ultracold plasma} 

The nitric oxide molecular ultracold plasma contains a persistent residue of NO Rydberg molecules that retain the initially selected principal quantum number, $n_0$.  The application of a weak radio frequency field subtly changes the state of this plasma in a way that causes the predissociation of this residue to accelerate.  

The NO Rydberg residue represents a surviving fraction of $n_0$ molecules trapped in states of high $\ell$ populated during the avalanche and quenched to a regime of suppressed $\ell$-mixing.  Predissociation resumes when the rf field acts to scramble the orbital angular momentum of those shelved, high-$\ell$, $n_0$ Rydberg molecules.  

The fractional yield of $n_0$ Rydberg molecules undergoing radio frequency field-accelerated predissociation varies substantially with plasma density.  This excludes $\ell$-mixing as a field-induced process in isolated-molecules.  We know that charged particles in a plasma also respond collectively to a radio frequency field by executing oscillatory modes of electron and ion motion.  Damping via charge coupling and collisions couples energy from the rf field to the plasma.  If this heating exceeds a mobility threshold for localized electrons, we should expect to see an effect of the rf field in an increase of the frequency of electron Rydberg collisions.  

That appears to be the case here.  The kinetics tell us that the rf field acts on the plasma to mobilize electrons.  These electrons collide with Rydberg molecules and scramble $\ell$.  Thus, we see an effect in the arrested residue of $n_0$ Rydberg molecules owing to a process that occurs in the ultracold plasma background.  Note that this mobilization, which dramatically accelerates predissociation kinetics, causes little change in the ultracold plasma background.  

We can therefore regard this adventitious population of high-$\ell$ $n_0$ Rydberg molecules and its response to an rf field as a quantum-state probe of the dynamics of avalanche and quench that form the molecular ultacold plasma.  The natural state of arrested predissociation points to an immobility of electrons in the quenched plasma.  A radio frequency field acts to mobilize electrons, which increases the rate of predissociation in a fraction of $n_0$ Rydberg population that grows with the density of the background plasma.  

Important questions remain.  If we can conclude that a radio frequency field promotes electron-Rydberg collisions, what initial state of the ultracold plasma serves as the source of electrons mobilized by the rf field?   What constrains the mobility of these electrons under field-free conditions?  How do we describe the interaction with the rf field that causes this increase in electron mobility?

\begin{acknowledgements}

{\bf Funding:}  This work was supported by the US Air Force Office of Scientific Research (Grant No. FA9550-17-1-0343), together with the Natural Sciences and Engineering research Council of Canada (NSERC), the Canada Foundation for Innovation (CFI) and the British Columbia Knowledge Development Fund (BCKDF).  JS gratefully acknowledges support from the National Science Foundation (NSF) Materials Research Science and Engineering Centers (MRSEC) program through Columbia University in the Center for Precision Assembly of Superstratic and Superatomic Solids under Grant No. DMR-1420634.  {\bf Author contributions:} R.W., M.A., F.B.V.M., J.S.K. and E.R.G designed the experiment.  R.W., M.A., F.B.V.M., performed the measurements.  R.W., M.A., K.L.M., K.M.G., J. S., F.B.V.M., J.S.K. and E.R.G analyzed the results and performed model calculations.  R.W., M.A. and E.R.G. wrote the manuscript with input from all authors.  {\bf Competing interests:} The authors declare that they have no competing interests. {\bf Data and materials availability:}  All data needed to evaluate the conclusions in the paper are present in the paper.  Additional data related to this paper may be requested from the authors.

\end{acknowledgements}

\bibliography{RF}% Produces the bibliography via BibTeX.

\end{document}